\pdfoutput=1

\documentclass[final,3p,times,review]{elsarticle}

\usepackage{amsmath,amssymb}
\usepackage{multirow}
\usepackage{lineno,hyperref}
\usepackage{array,tabularx}
\usepackage{graphicx}
\usepackage{footmisc} 
\usepackage{verbatim}
\usepackage{geometry}
\usepackage{subfigure}
\usepackage{caption}
\usepackage{lineno}
\usepackage{color}
\usepackage{bm}
\usepackage{booktabs}

\biboptions{numbers,sort&compress}

\begin{document}

\begin{frontmatter}
\title{Unfolding environmental $\gamma$ flux spectrum with portable CZT detector\tnotemark[mytitle]}
\tnotetext[mytitle]{Supported by National Key R\&D Program of China (2023YFA1607203), National Natural Science Foundation of China (12005225, 12141505) and the Fundamental Research Funds for the Central Universities, China (WK2360000015)
} 

\author[1,2]{Taiyuan Liu}
\author[1,2]{Mingxuan Xue\corref{mycorrespondingauthor}}
\cortext[mycorrespondingauthor]{Corresponding author}
\ead{xuemx@ustc.edu.cn}
\author[1,2]{Haiping Peng}
\author[1,2]{Kangkang Zhao}
\author[1,2]{Deyong Duan}
\author[1,2]{Yichao Wang}
\author[1,2]{Changqing Feng}
\author[1,2]{Yifeng Wei}
\author[1,2]{Qing Lin}
\author[1,2]{Zizong Xu}
\author[1,2]{Xiaolian Wang}

\address[1]{{State Key Laboratory of Particle Detection and Electronics, University of Science and Technology of China},
            {Hefei},
            {230026}, 
            {China}}
\address[2]{{Department of Modern Physics, University of Science and Technology of China},
            {Hefei},
            {230026}, 
            {China}}

\begin{abstract}
Environmental $\gamma$-rays constitute a crucial source of background in various nuclear, particle and quantum physics experiments. To evaluate the flux rate and the spectrum of $\gamma$ background, we have developed a novel and straightforward approach to reconstruct the environmental $\gamma$ flux spectrum by applying a portable CZT $\gamma$ detector and iterative Bayesian unfolding, which possesses excellent transferability for broader applications. In this paper, the calibration and GEANT4 Monte-Carlo modeling of the CZT detector,  the unfolding procedure as well as the uncertainty estimation are demonstrated in detail. The reconstructed spectrum reveals an environmental $\gamma$ flux intensity of $3.3\pm 0.9\times 10^{7}$~ (m$^2\cdot$sr$\cdot$hour)$^{-1}$ ranging from 73 to 3033~keV, along with characteristic peaks primarily arising from $^{232}$Th series, $^{238}$U series and $^{40}$K. We also give an instance of background rate evaluation with the unfolded spectrum for validation of the approach. 

\end{abstract}


\end{frontmatter}

\section{Introduction}

Evaluation and reduction of backgrounds, including environmental $\gamma$-rays and neutrons, cosmic rays and material radioactivities are the main issues to be addressed for particle physics experiments searching for rare events such as dark matter particle interactions and neutrinoless double beta decay~\cite{CUORE:2021mvw, augier2022final, ahmine2023test, alkhatib2021light}, biological and quantum physics experiments requiring low background environments~\cite{thome2017repair, Vepsalainen:2020trd}. To minimize the adverse impact of background events, evaluating and reducing environmental $\gamma$-rays is crucial , especially in above-ground labs for a wide range of radioactive sources and large background intensity.

A common approach to reduce environmental $\gamma$ background is to set up shielding materials, such as low-background lead and copper. To facilitate the design of shielding, simulations of $\gamma$ background that are widely adopted currently require the detailed information of both the spatial and activity distributions of radioactive elements\cite{CUORE:2017ztm}, which can be challenging to acquire precisely in some laboratories. Therefore, we have developed a novel and highly feasible approach to obtain the $\gamma$ spectra and its flux intensity in local space near the detector utilizing a Cadmium Zinc Telluride (CdZnTe, CZT) detector and iterative Bayesian unfolding algorithm. Simulations based on the obtained spectrum can be implemented to estimate the $\gamma$ background  with acceptable precision for our experiments.

This paper is organized as below: Section~\ref{sec:calibration} presents the calibration and peak parameterization of the CZT detector, Section~\ref{sec:CZTmodel} shows the GEANT4 modeling of the detector to scrutinize its response to $\gamma$-rays, Section~\ref{sec:deconvolution} elaborates the procedure of $\gamma$ flux spectrum unfolding, and Section~\ref{sec:design} demonstrates a comparison of background spectra as well as total counting rates between experimental measurement and Monte-Carlo (MC) simulations based on the unfolded spectrum.

\section{Calibration of CZT $\gamma$-detector}\label{sec:calibration}

\subsection{$\gamma$-rays spectrum measurement of radioactive sources}

\begin{figure}[!htbp]
\begin{center}
\subfigure[]{
\includegraphics
  [width=5cm]
  {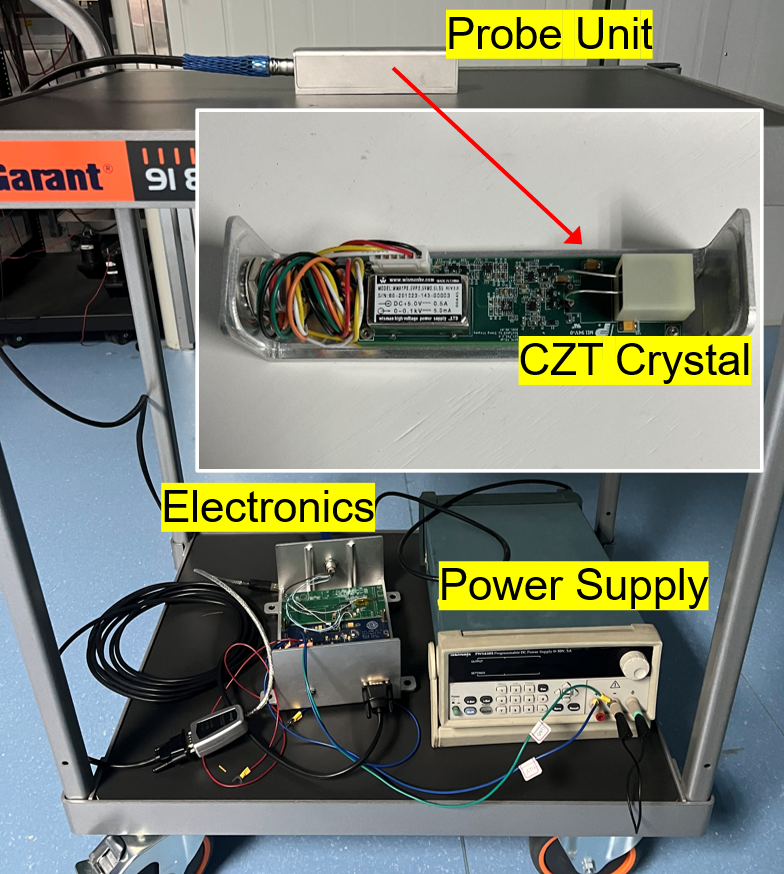}}
\subfigure[]{
\includegraphics
  [width=8.5cm]
  {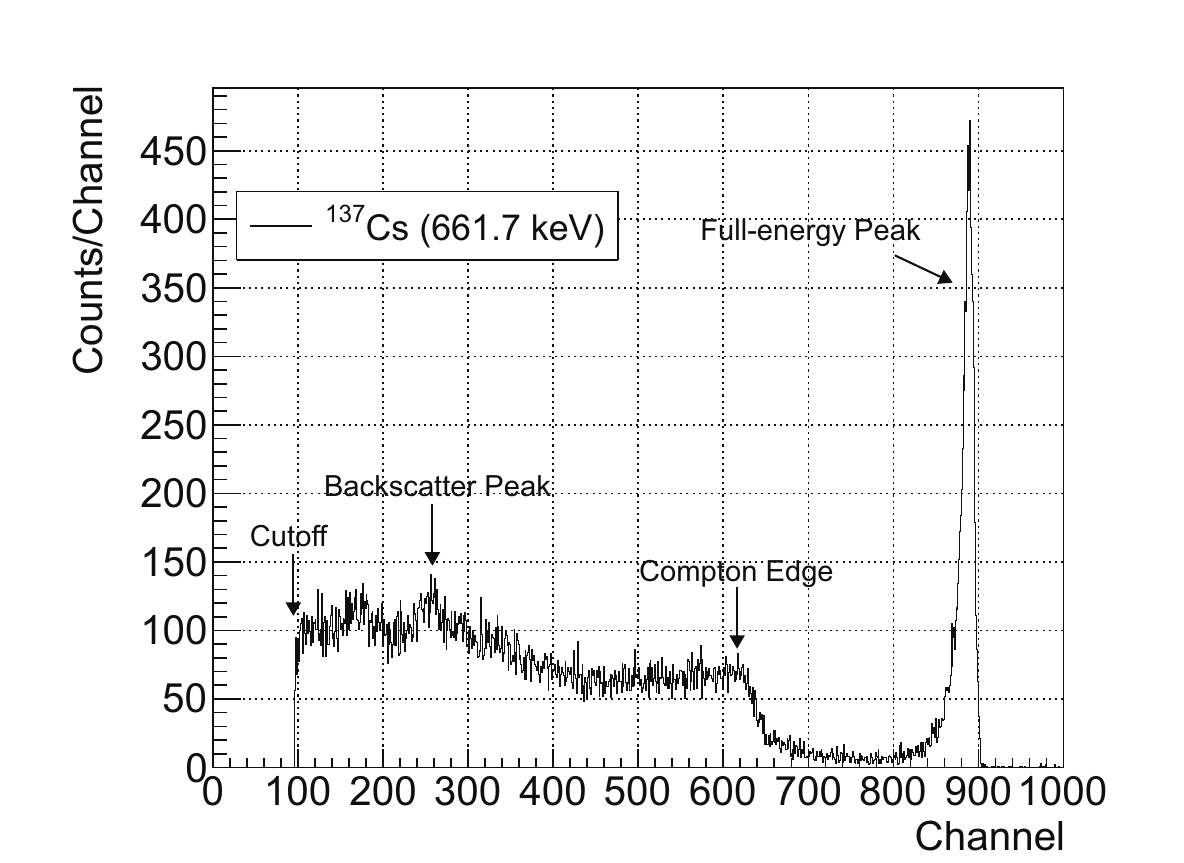}}
\caption{\label{fig:Cs137SpectrumANDapparatus}(a) CZT detector  employed in this study. The size of the crystal is 10 mm $\times$ 10 mm $\times$ 5 mm.  (b) Spectrum of a $^{137}$Cs source placed measured with CZT detector.}
\end{center}
\label{fig:CZTEfield}
\end{figure}

CZT detectors have several unique advantages in $\gamma$-ray spectrometry, including high energy resolution, ability to work at room temperature, and excellent portability, thus making them an ideal option for environmental $\gamma$ background measurement. However, the drawbacks of CZT detectors are non-negligible: the small crystal volume yields low efficiency, and incomplete charge collection within the CZT crystal leads to non-Gaussian detector responses, noticed by the asymmetric shape of the full-energy peaks. The detector used in this study, as shown in Fig.~\ref{fig:Cs137SpectrumANDapparatus}~(a), has 4096 ADC channels with a full measurement range of approximately 3000~keV. A low-energy cutoff is set at channel 96 out of 4096 due to the ADC amplitude threshold.

For the calibration of the CZT detector, several radioactive sources are utilized, including $^{137}$Cs, $^{22}$Na, $^{60}$Co and $^{208}$Tl, as summarized in Table~\ref{tab:sources}. Figure~\ref{fig:Cs137SpectrumANDapparatus}~(b) shows the $\gamma$ spectrum of a $^{137}$Cs source positioned in front of the CZT detector for a duration of 10 minutes. Several features of $\gamma$ spectrum are observed obviously, such as full-energy peak, backscattering peak and Compton plateau. Notably, the full-energy peak, corresponding to an energy deposition of 661.7~keV, is located around channel 880 with an energy resolution of $1.8\%$ (FWHM, Full Width at Half Maximum). Meanwhile a prominent low-energy tailing effect, which leads to a deviation from Gaussian distribution, is also observed clearly.
The reasons and solutions to describe this tail are considered in the following section.

\begin{table}[!htb]
\caption{\label{tab:sources}Radioactive sources used for calibration. To avoid background from Compton plateaus, only the full-energy peaks with highest energies from $^{22}$Na, $^{60}$Co and $^{208}$Tl are used.}
\begin{center}
\begin{tabular*}{5cm} {@{\extracolsep{\fill} } cc}
\toprule
Sources & Energy (keV)  \\
\midrule
$^{137}$Cs  & 661.7   \\
$^{22}$Na   & 1274.5      \\
$^{60}$Co   & 1332.5   \\
$^{208}$Tl  & 2614.5   \\
\bottomrule
\end{tabular*}
\end{center}
\end{table}

\subsection{Fitting of the full-energy peaks}

Typically, CZT detectors have poor hole charge collection properties due to crystal defects~\cite{eisen1998cdte}, which manifests itself mainly in the following ways: (a) the full-energy peaks deviate from Gaussian function; (b) low-energy tails appear beneath the full-energy peaks. Several parameterization methods have been developed to describe the full-energy peaks~\cite{mortreau2001characterisation,namboodiri1996gamma,dardenne1999cadmium}. In this study, the full-energy peak is fitted with a combination of a Gaussian function and an exponential function which minimizes the number of parameters while maintaining accurate description of the peaks. 
The fitting function can be expressed as

\begin{equation}\label{eq:peakfunction}
\begin{aligned}
N(C _i)=N_0[e^{-(C _i - C _0)^2/(\sqrt{2}\sigma )^2}+T(C _i)]+Bg(C _i),
\end{aligned}
\end{equation}

\noindent where $N(C _i)$ denotes the counts in bin $C _i$, $Bg(C _i)$ represents the background function where the complementary error function (Erfc) is used mainly for background from multiple Compton scattering, and $T(C _i)$ represents the tail function

\begin{equation}\label{eq:tailfunction}
\begin{aligned}
T(C _i)=Ae^{B(C _i - C _0)}[1-e^{-(C _i - C _0)^2/(\sqrt{2}\sigma )^2}]\Theta(C _i - C _0),
\end{aligned}
\end{equation}

\noindent where $\Theta(C _i - C _0)$ is the step function which equals $1$ for $C _i < C _0$, and $0$ for $C _i > C _0$. There are five parameters in the fitting function characterizing a full-energy peak: 
$N _0$ is the counts at the peak, $C _0$ and $\sigma$ are mean and standard deviation of the Gaussian function, meanwhile $A$ and $B$ represent the amplitude and slope of the tail, respectively. 

\begin{figure}[!htbp]
\begin{center}
\includegraphics
  [width=8.5cm]
  {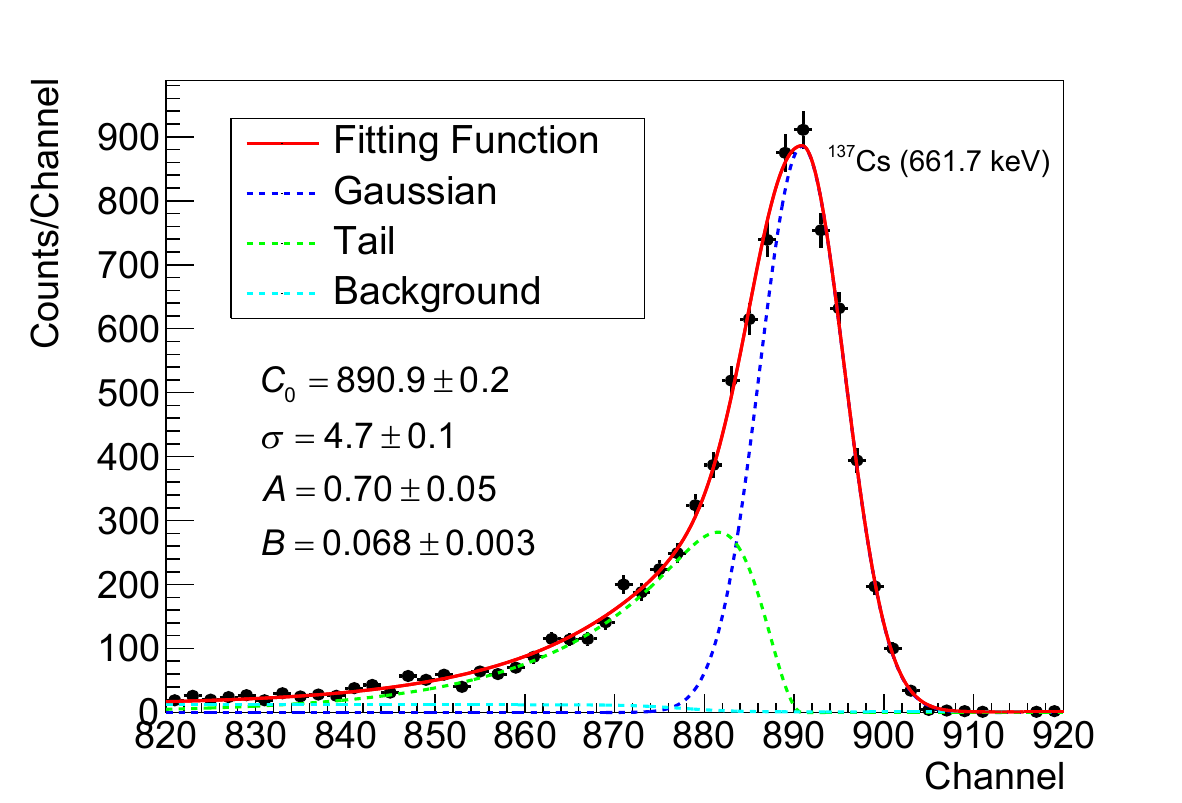}
\caption{\label{fig:peakfit}A five-parameter fit to the $661.7$~keV full-energy peak from $^{137}$Cs source.}
\end{center}
\end{figure}

The fitting is made with ROOT~\cite{Brun:1997pa} v6.26/06 applying the log likelihood method. Figure~\ref{fig:peakfit} shows the fitting components of $^{137}$Cs full-energy peak with energy $661.7$~keV. Similar fittings are performed on the full-energy peaks of source listed in Table~\ref{tab:sources} to obtain the energy dependences of the parameters.

\begin{figure}[!htbp]
\begin{center}
\includegraphics
  [height=6cm]
  {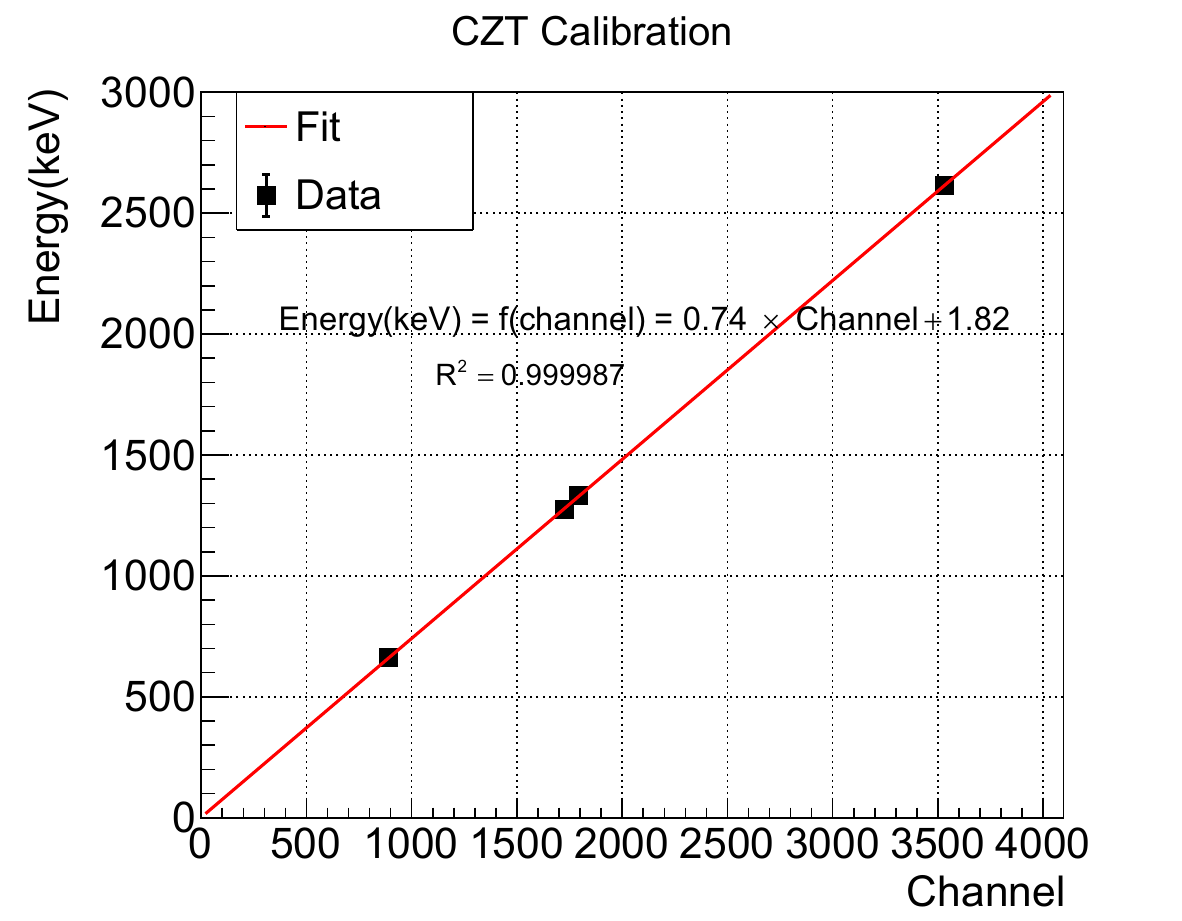}
\caption{\label{fig:parEnergyDependencea}Energy calibration of the CZT detector.}
\end{center}
\end{figure}

\begin{figure}[!htbp]
\begin{center}
\includegraphics
  [height=6cm]
  {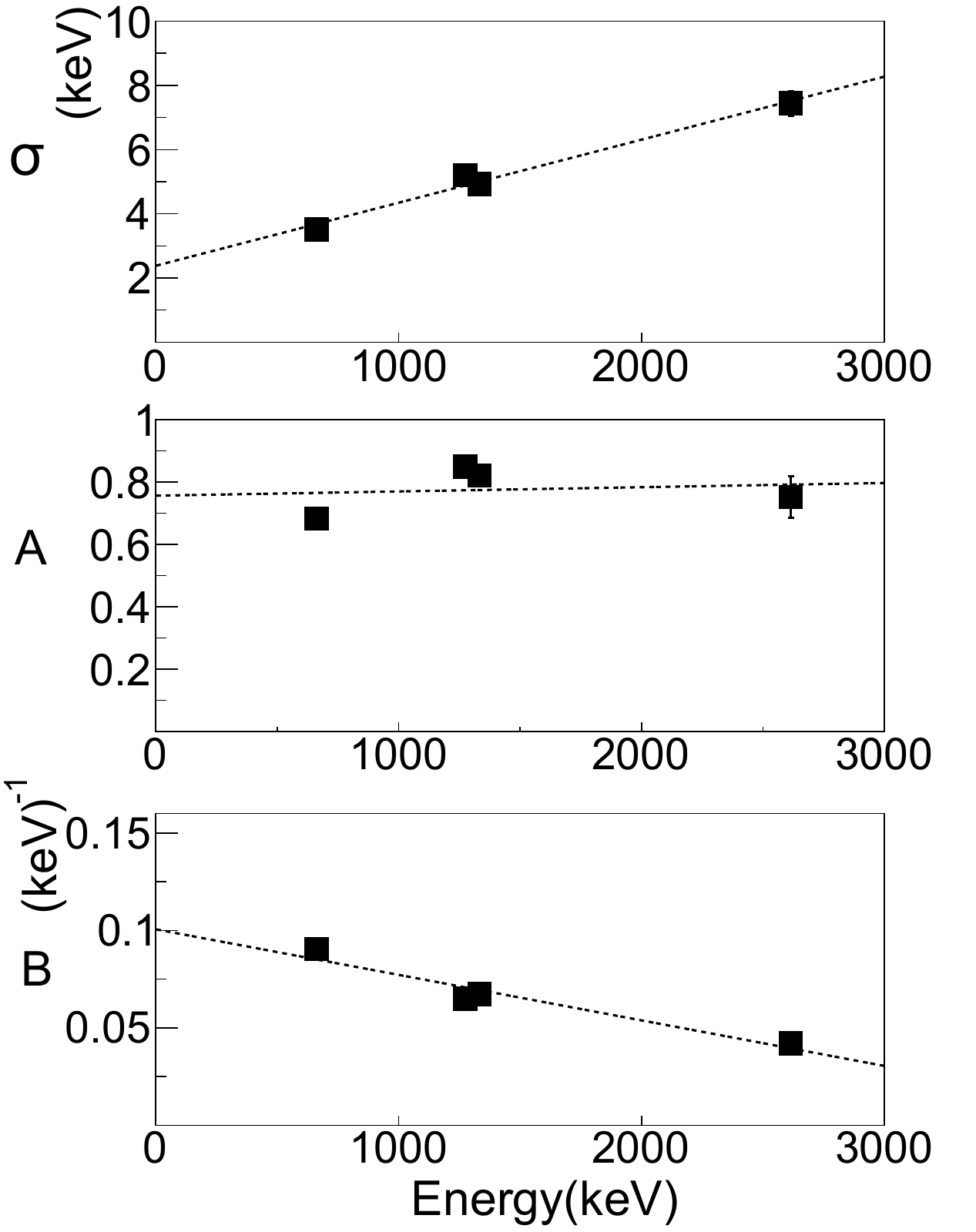}
\caption{\label{fig:parEnergyDependenceb}Energy dependences of the parameters. 
}
\end{center}
\end{figure}

For energy calibration, the mean value of Gaussian function $C _0$ is regarded as the specific channel where the full energy peak locates. A linear fitting is performed on $C _0$ as a function of corresponding energy, as shown in Fig.~\ref{fig:parEnergyDependencea}. The result illustrates excellent linearity with $R^2 = 0.999987$ in the given energy range, from which the cutoff channel 96 is estimated to be 73~keV and the maximal channel 4096 to be 3033~keV. Meanwhile, the energy dependences of the parameters $A$, $B$, $\sigma$ are also derived from peak fitting, as shown in Fig.~\ref{fig:parEnergyDependenceb}, where the standard deviation $\sigma$ tends to increase with energy, the tail slope $B$ tends to descend with energy, and the tail amplitude $A$ remains relatively stable. To obtain continuous energy-dependent functions in a feasible way, the linear fits are also implemented based on the available data points, as shown in Fig.~\ref{fig:parEnergyDependenceb}. 
The energy-dependent parameters are used to reconstruct the low-energy tailing effect and energy resolution of the CZT detector in the following Monte-Carlo simulations by smearing the energy depositions.

\section{Monte-Carlo simulation of CZT detector}\label{sec:CZTmodel}

\subsection{Detector Modeling}

Deconvolution and reconstruction of the environmental $\gamma$ flux spectrum requires a thorough understanding of the CZT detector’s response to the incident 
$\gamma$-rays. In addition to the resolution effect elaborated in Section~\ref{sec:calibration}, a Monte-Carlo simulation is conducted with GEANT4~\cite{GEANT4:2002zbu} v11.0.3 toolkit to study the related physical processes of the $\gamma$-rays interacting with the CZT detector.

\begin{figure}[!htbp]
\begin{center}
\includegraphics
  [width=8.5cm]
  {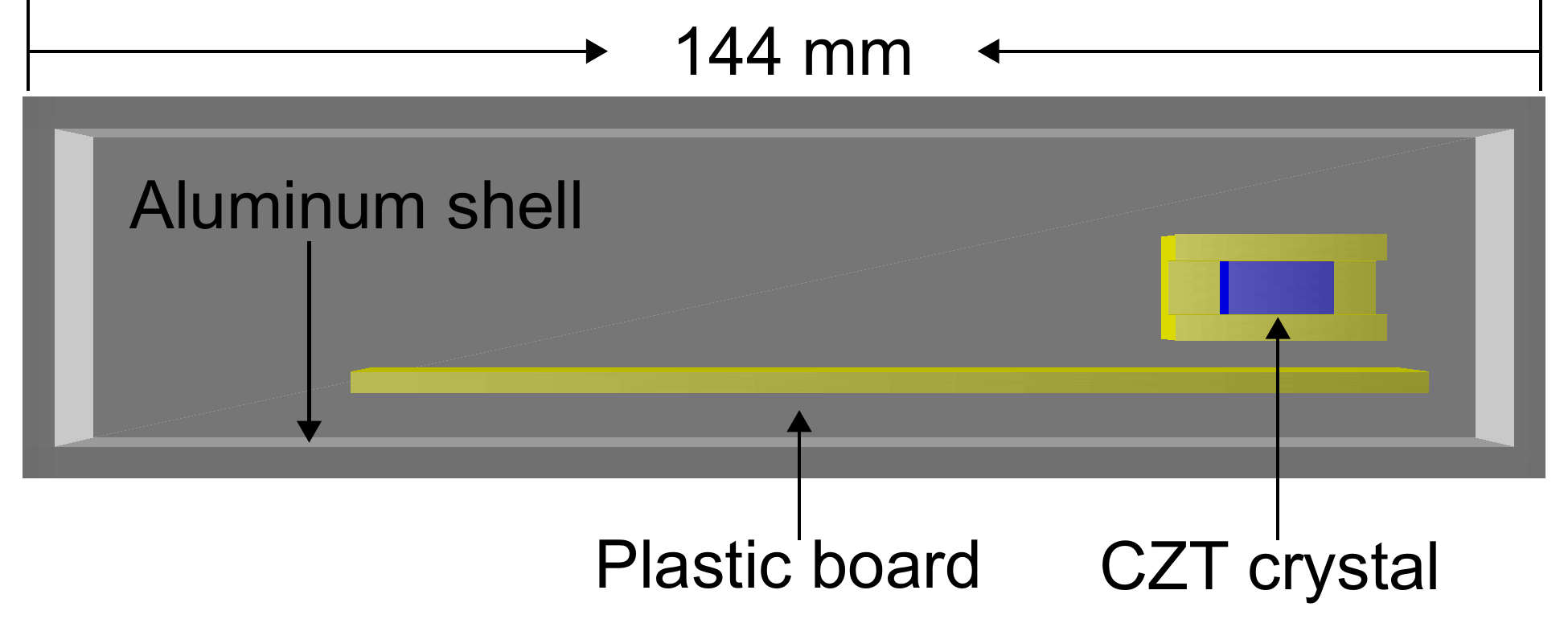}
\caption{\label{fig:CZTgeant4}A sketch of GEANT4 geometry of the detector. The gray part is the 3 mm-thick aluminum shell. The yellow part is the plastic board and the holder. The blue part is the $10\ \mbox{mm}\times 10\ \mbox{mm}\times 5\ \mbox{mm}$ CZT crystal.}
\end{center}
\end{figure}

\begin{figure}[htbp]
\begin{center}
\includegraphics
  [width=8.5cm]
  {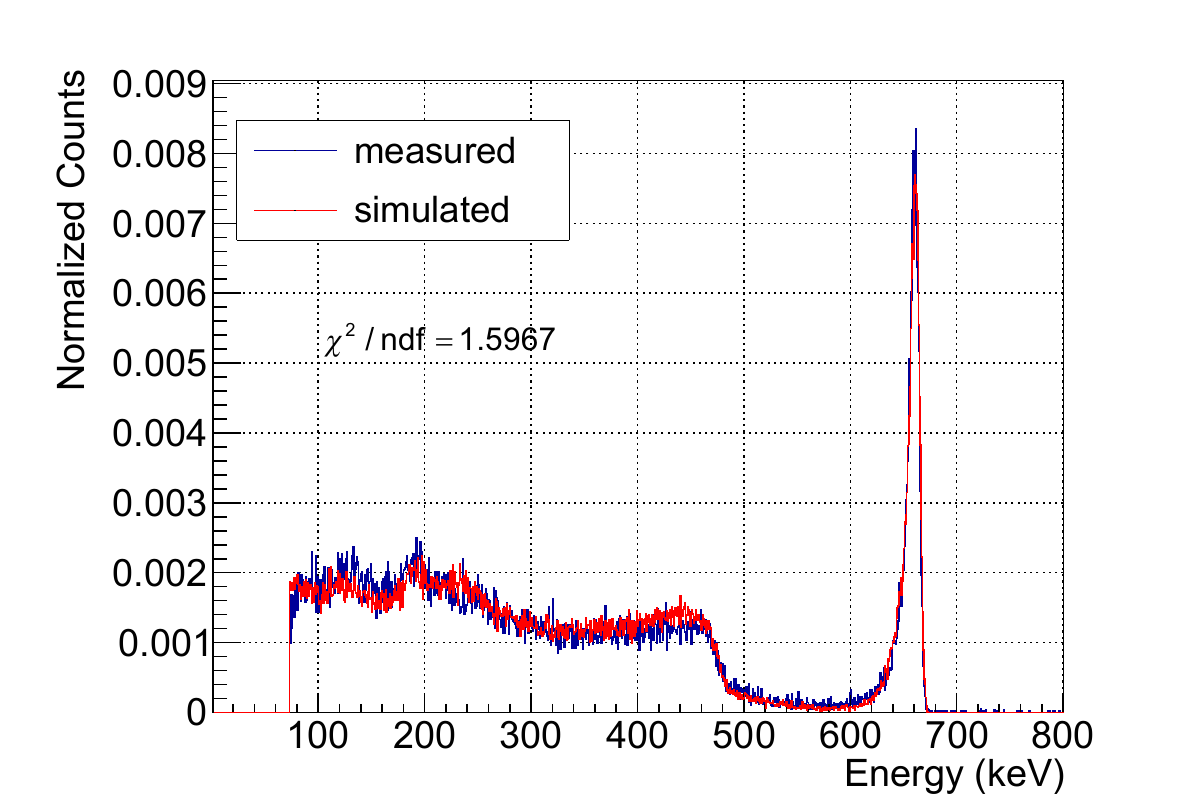}
\caption{\label{fig:CZTmodelverify}Comparison between measured and simulated spectrum of  $^{137}$Cs source.}
\end{center}
\end{figure}

Figure~\ref{fig:CZTgeant4} depicts the detector geometry defined in GEANT4. Electromagnetic processes of photons, electrons and positrons~\cite{geant4EM,geant4EM2} are registered to describe the interactions of $\gamma$-rays in the CZT crystal, and G4RadioactiveDecay module is used to describe the decays of $\gamma$ sources. The physical events of both radioactive nuclides and $\gamma$-rays are generated with G4GeneralParticleSource. In the simulation, the energy deposition of $\gamma$-rays in the CZT crystal is traced and logged until they either get absorbed or escape from the crystal. Subsequently, the deposited energies are smeared according to the peak fitting function described in section \ref{sec:calibration} to obtain the final simulated energy spectrum. To assess the accuracy of modeling, a $^{137}$Cs source is set up in the MC simulation in accordance with the experimental configuration for comparison. As shown in Fig.~\ref{fig:CZTmodelverify}, a good consistency between the measured and simulated energy spectra, which is normalized by counts, can be observed. In addition, according to the efficiency calibrations of semiconductor detectors conducted in some previous researches\cite{liu2017detection, joel2018monte}, Geant4 is able to accurately describe the electromagnetic processes of $\gamma$-rays in our interested energy region. Therefore, the model is supposed to be reliable in calculating the response of the CZT detector.

\subsection{Detector response}

\begin{figure}[htbp]
\begin{center}
\includegraphics
  [width=6.5cm]
  {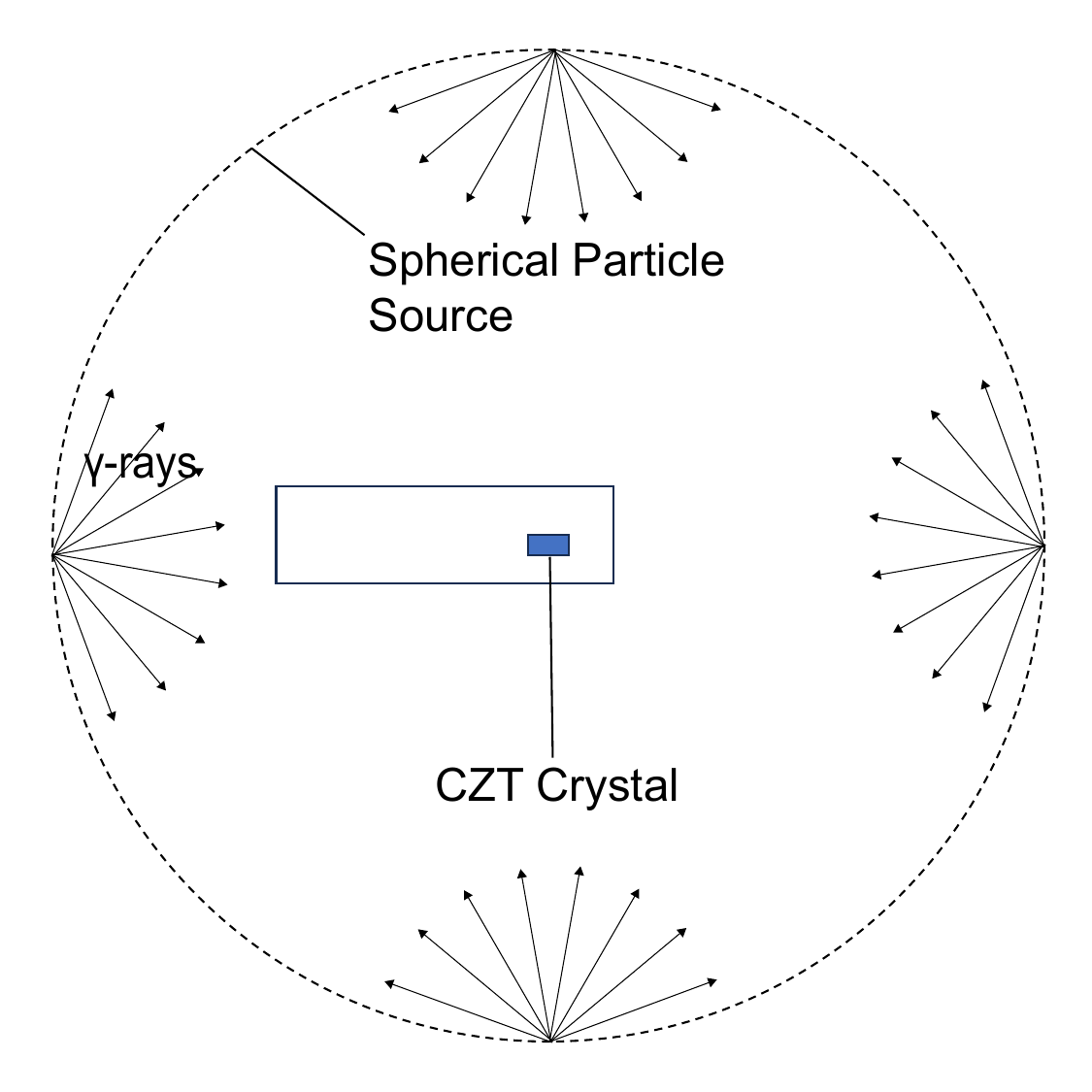}
\caption{\label{fig:CZTsimu}Setup of detector and particle source in simulation.}
\end{center}
\end{figure}

With MC simulations, we can scrutinize the response of the CZT detector to $\gamma$-rays with different energy. Figure~\ref{fig:CZTsimu} schematically shows the MC simulation setup, wherein $\gamma$-rays are assumed to emit from a spherical particle source enveloping the CZT detector~\cite{Niu_2015}, with energy  distributed uniformly from 73 to 3033~keV and directions following Lambert's cosine law to simulate isotropic radiation in space. 
$10^{10}$ events are generated in total.

For the $\gamma$-rays with the initial energy $E_{true}=E^i$, we can define the efficiency $\varepsilon _i$, which represents the joint effects of geometric acceptance and intrinsic efficiency, and can be expressed as

\begin{equation}\label{eq:unfolding3}
\varepsilon _i = \frac{N(E_{obs}>E_{cut}~\&~ E_{true}=E^i)}{N(E_{true}=E^i)},
\end{equation}

\noindent where $i$ is the bin index, $E_{obs}$ denotes the observed energy deposition in the measured spectrum, $E_{cut}$ is the low-energy cutoff explained in Section. \ref{sec:calibration}, $N(E_{true}=E^i)$ denotes the total number of $\gamma$-rays with $E_{true}=E^i$ emitted from the spherical source, and $N(E_{obs}>E_{cut}~\&~ E_{true}=E^i)$ is a part of $N(E_{true}=E^i)$ that leaves observable energy deposition.


Due to the different physical processes and noises in the CZT detector, for the $\gamma$-rays of the same initial energy, the observed energy deposition can also distribute broadly. The migration matrix $R$ indicates such distributions, with each element $R_{ij}$ defined by

\begin{equation}\label{eq:unfolding1}
R_{ij} = P(E_{obs}=E^j~|~E_{true}=E^i) = \frac{N(E_{obs}=E^j~\&~ E_{true}=E^i)}{N(E_{true}=E^i)},
\end{equation}

\noindent where $P(E_{obs}=E^j|E_{true}=E^i)$ represents the conditional probability that a $\gamma$-ray with given energy $E_{true}=E^i$ produces an event with $E_{obs}=E^j$. 

\begin{figure}[!htbp]
\begin{center}
\subfigure[]{
\includegraphics
  [height=5cm]
  {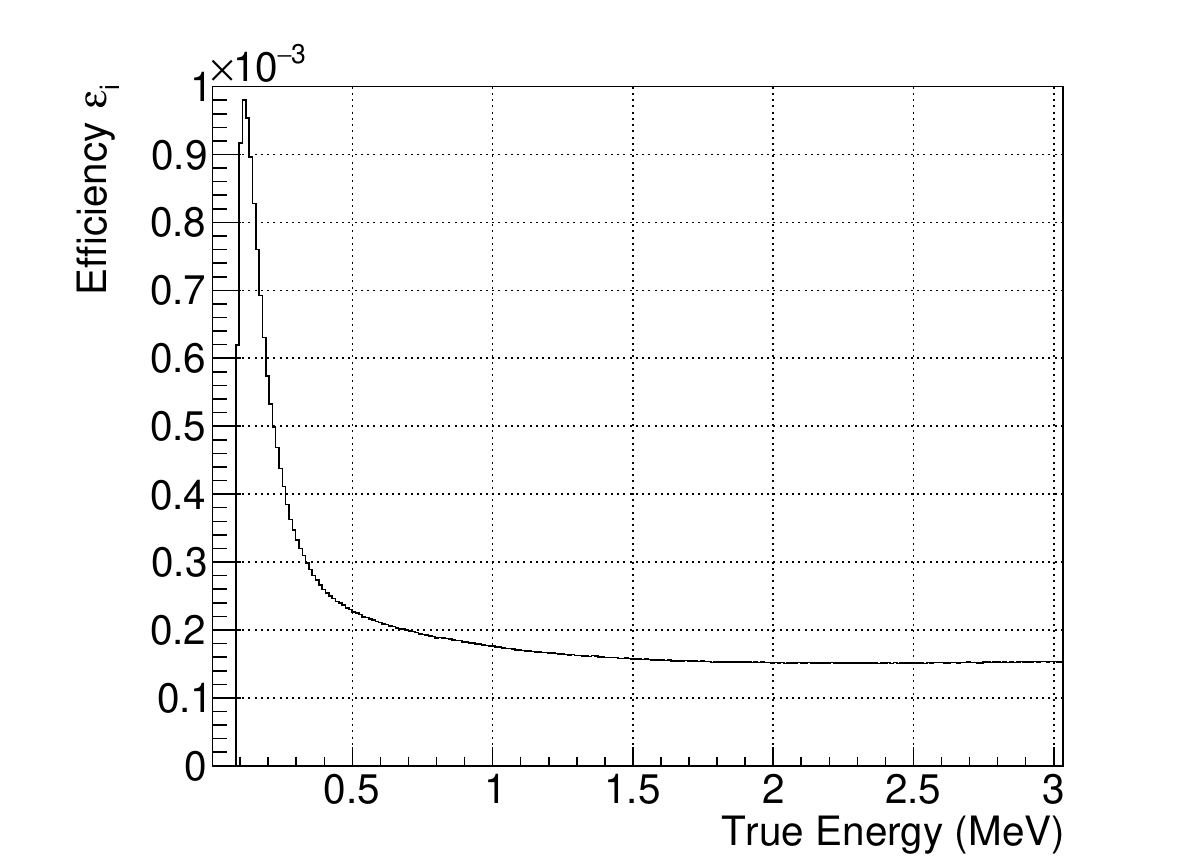}}
\subfigure[]{
\includegraphics
  [height=5cm]
  {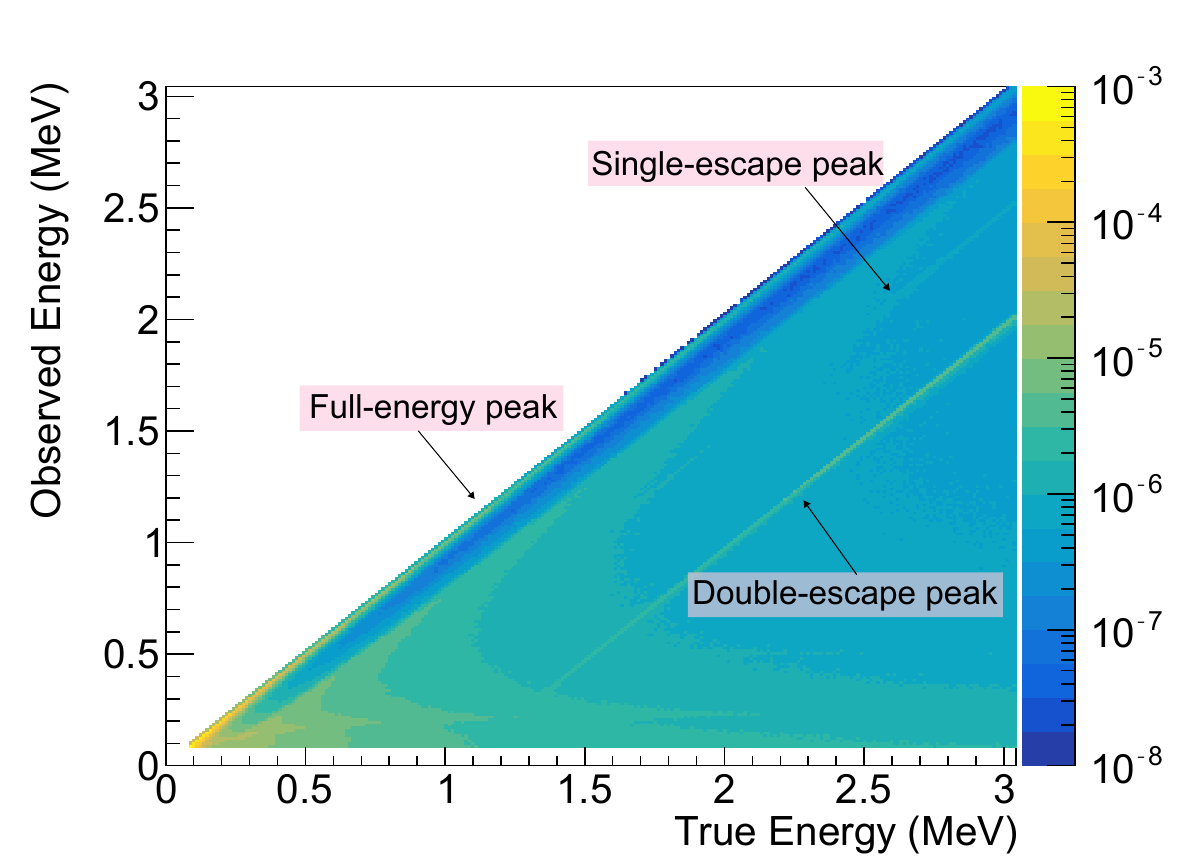}}

\caption{\label{fig:CZTResponse}(a)Joint effects of detector geometry acceptance and intrinsic efficiency $\varepsilon _i $, (b) Migration matrix R of the CZT detector.}
\end{center}
\end{figure}

The efficiency curve $\varepsilon _i $, as well as the migration matrix $R$ of the CZT detector are derived from the MC simulations for spectrum deconvolution, as shown in Fig.~\ref{fig:CZTResponse}~(a) and (b). $\varepsilon _i$ decreases with energy due to the total interaction cross section of photons reduces as the energy increases. And the migration matrix reveals several features of general $\gamma$ spectrum: the diagonal band represents the full-energy peaks, the bands on the lines $E_{obs} = E_{true} - 0.511$~MeV and $E_{obs} = E_{true} - 1.022$~MeV represent the single and double escape peaks respectively, and the others are dominated by Compton scattering.

\section{Deconvolution of $\gamma$ spectrum}\label{sec:deconvolution}

\subsection{Data taking}

\begin{figure}[!htbp]
\begin{center}
\includegraphics
  [width=8cm]
  {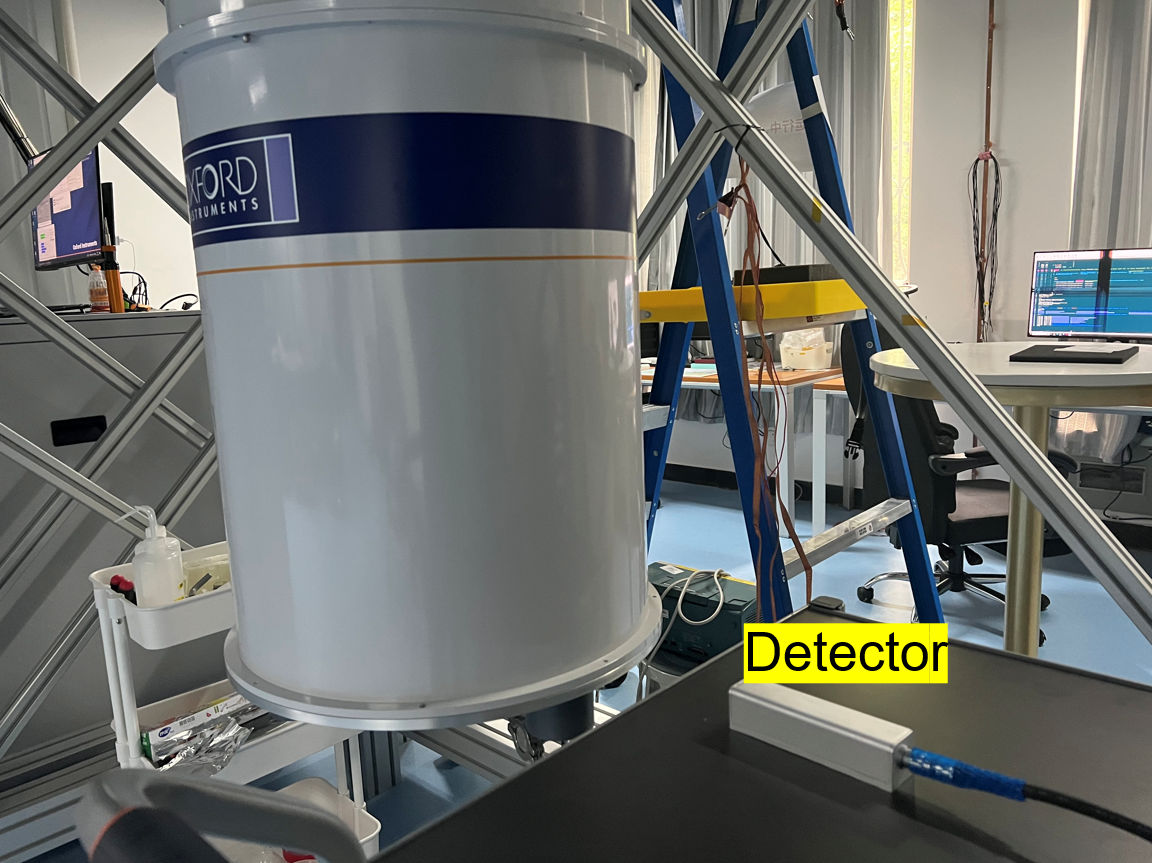}
\caption{\label{fig:dataking_setup}Experimental setup at Platform for Cryogenic Detector R\&D.}
\end{center}
\end{figure}

\begin{figure}[!htbp]
\begin{center}
\includegraphics
  [width=12cm]
  {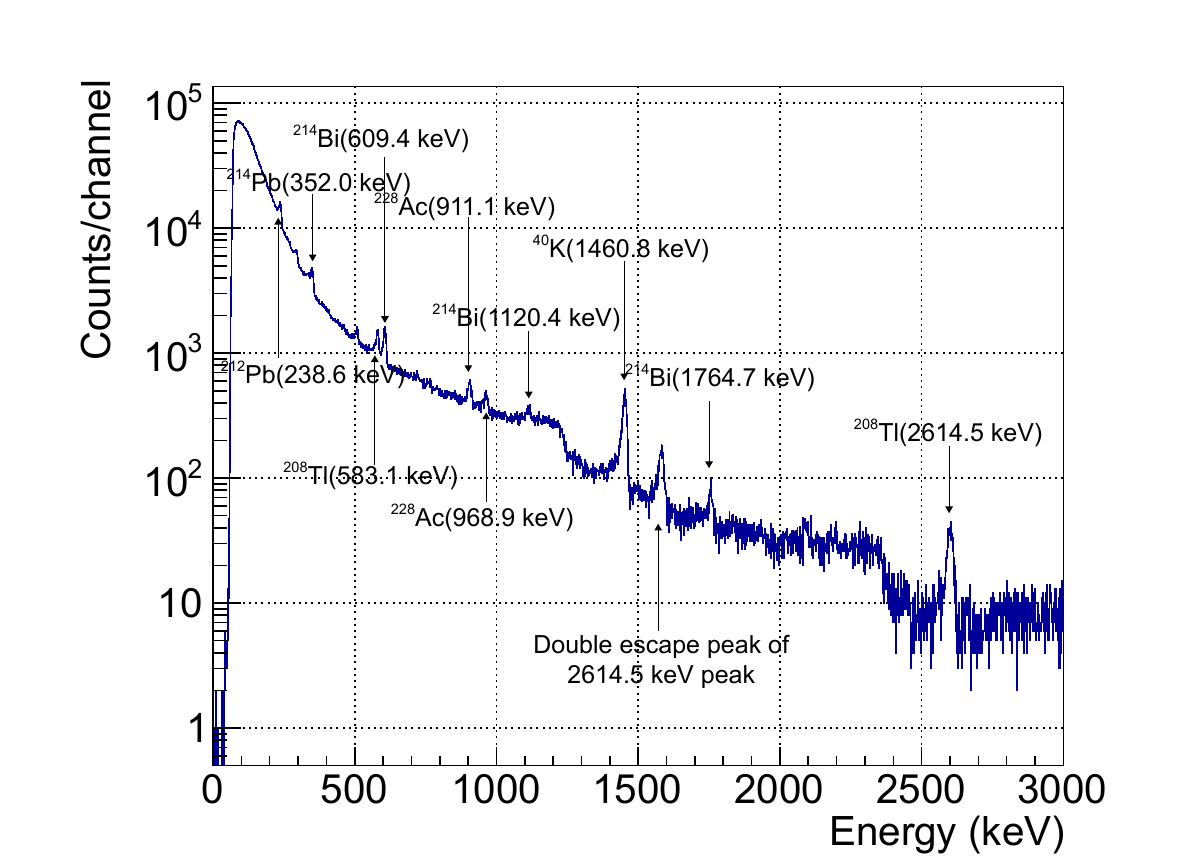}
\caption{\label{fig:dataking}Environmental $\gamma$ spectrum measured with CZT detector for 327 hours.}
\end{center}
\end{figure}

Platform for Cryogenic Detector R\&D, located at an above-ground laboratory at University of Science and Technology, is aimed at bolometer R\&D to search for neutrinoless double beta decay and requires background reduction. To obtain environmental $\gamma$ background spectrum and the flux around the platform, the CZT detector is positioned near the cryostat for measurement, as shown in Fig.~\ref{fig:dataking_setup}. The measured spectrum with an 
exposure time of 327 hours is shown in Fig.~\ref{fig:dataking}. Most of the peaks correspond to radioactive nuclides, such as those positioned at 609.4, 1460.8, 2614.5~keV. The deconvolution on the measured spectrum can be carried out to eliminate the detector's response effect and get the real $\gamma$ spectrum. Due to the precision limit of peak parameterization and the computing time consumption, the measured spectrum is rebinned from 4096 to 256 bins in the unfolding implementation.

\subsection{Unfolding procedure}

Iterative Bayesian unfolding~\cite{DAgostini:1994fjx} is carried out in this study for deconvolution. The procedure is generalized as below:

\begin{equation}\label{eq:unfolding5}
\begin{aligned}
N_{k+1}(E_{true}=E^i)=\frac{1}{\varepsilon _i}\sum\limits _{j=1}^{n_B} N(E_{obs}=E^j)P_k(E_{true}=E^i|E_{obs}=E^j),
\end{aligned}
\end{equation}

\noindent where $N_{k+1}(E_{true}=E^i)$ is the counts in the $i$-th bin of the reconstructed spectrum after $k+1$ iterations, $N(E_{obs}=E^j)$ is the counts in the $j$-th bin of the measured spectrum as shown in Fig.~\ref{fig:dataking}, $\varepsilon _i$ is the efficiency described in section~\ref{sec:CZTmodel}, $n_B = 256$ is the total number of bins, and $P_k(E_{true}=E^i|E_{obs}=E^j)$ represents the estimated conditional probability in the $k$-th iteration that an event with specific energy deposition $E_{obs}=E^j$ is produced by a $\gamma$-ray incident with $E_{true}=E^i$, which is estimated by the Bayes formula:

\begin{equation}\label{eq:unfolding6}
\begin{aligned}
P_k(E_{true}=E^i|E_{obs}=E^j)=\frac{P(E_{obs}=E^j|E_{true}=E^i)P_k(E_{true}=E^i)}{\sum ^{n_B}_{l=1}P(E_{obs}=E^j|E_{true}=E^l)P_k(E_{true}=E^l)},
\end{aligned}
\end{equation}

\noindent where $P(E_{obs}=E^j|E_{true}=E^i)$ is the element of the 
migration matrix $R_{ij}$ defined in Eq.~\ref{eq:unfolding1} and estimated in MC simulation; $P_k(E_{true}=E^i)$ represents the $i$-th bin of the normalized reconstructed spectrum in the $k$-th iteration. For initialization, $P_0(E_{true}=E^i)$ is set to a uniform distribution:

\begin{equation}\label{eq:unfolding8}
\begin{aligned}
P_0(E_{true}=E^i) = \frac{1}{n_{B}},
\end{aligned}
\end{equation}

From Eq. (\ref{eq:unfolding5}-\ref{eq:unfolding8}), an iterative algorithm for $\gamma$ spectrum unfolding is constructed. To terminate the iterations appropriately, we use a criterion based on $\chi^2$ comparison:

\begin{equation}\label{eq:unfolding9}
\begin{aligned}
\chi ^2 _{k} = \sum ^{n_B}_{i=1}\frac{(N_{k+1}(E_{true}=E^i)-N_k(E_{true}=E^i))^2}{N_k(E_{true}=E^i)},
\end{aligned}
\end{equation}

\noindent where $\chi ^2 _{k}$ measures the differences of two consecutive iterations. Too few iterations lead to a result far from convergence, while too many lead to larger propagated uncertainties and time consumption~\cite{DAgostini:1994fjx, Bueno:2022dva}. As $\chi ^2 _{k}$ continually decreases with the number of iterations, we choose to stop at iteration 500, where the difference is small enough ($\chi ^2 _{500} < 10^{-2}$). The unfolding program is based on RooUnfold~\cite{Brenner:2019lmf} v2.0.0.

\subsection{Flux calculation}

The deconvolution of measured spectrum gives the counting spectrum of $\gamma$-rays passing through the virtual source sphere with a radius of $r$, as Fig.~\ref{fig:CZTsimu} indicates. With the approximation that the background flux is isotropic in space, the counts can be converted to flux:

\begin{equation}\label{eq:flux}
\begin{aligned}
\varPhi(E_{true}=E^i) = \frac{N(E_{true}=E^i)}{T\cdot A \cdot \Delta E},
\end{aligned}
\end{equation}

\noindent where $\varPhi(E_{true}=E^i)$ is the flux of $\gamma$-rays with energy with $E_{true}=E^i$, $T$ is the exposure time, $\Delta E$ is the bin interval, and $A$ is the geometric acceptance of the sphere:

\begin{equation}
\begin{aligned}
A = \pi S_{source} = 4\pi ^2 r^2.
\end{aligned}
\label{eq:unfolding11}
\end{equation}

\subsection{Result of $\gamma$ flux spectrum unfolding}

\begin{figure}[!htbp]
\begin{center}
\includegraphics
  [width=15cm]
  {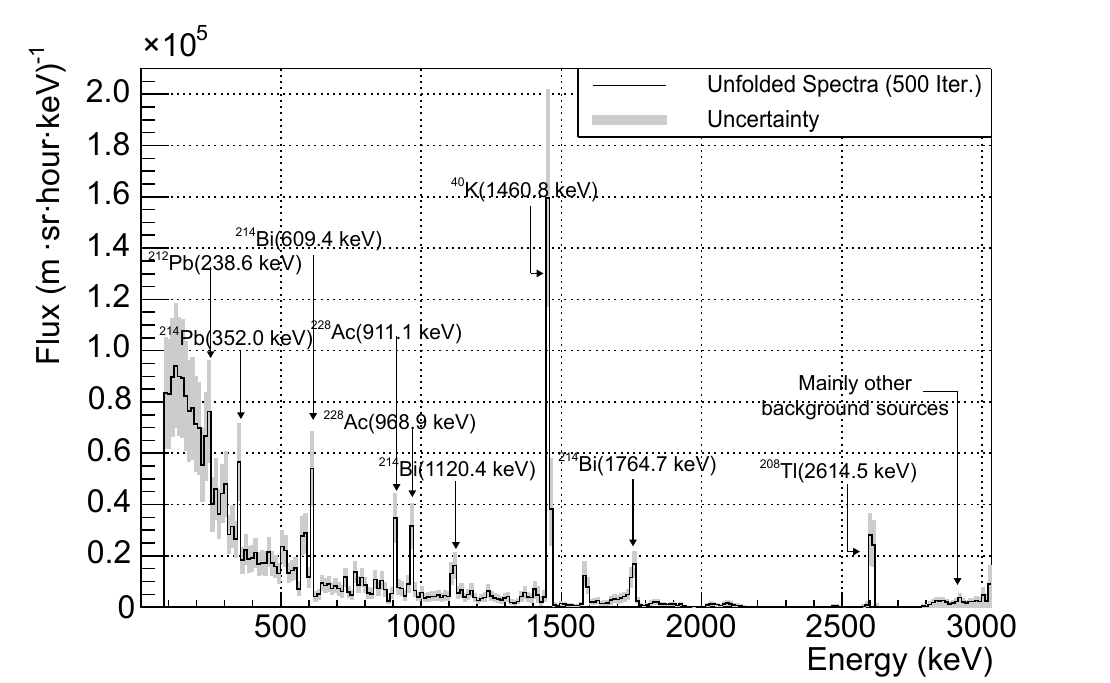}
\caption{\label{fig:unfoldedspectra}Reconstructed environmental $\gamma$ flux spectrum.}
\end{center}
\end{figure}

The environmental $\gamma$ flux spectrum deconvoluted with iterative Bayesian unfolding algorithm is illustrated in Fig.~\ref{fig:unfoldedspectra}. The spectrum reveals several characteristic peaks attributed to different radioactive nuclides: $^{228}$Ac, $^{212}$Pb and $^{208}$Tl which belong to $^{232}$Th nuclide series; $^{214}$Bi and $^{214}$Pb which belong to $^{238}$U nuclide series; as well as $^{40}$K itself. In addition, there is a continuous background predominantly distributed across low-energy regions\footnote{We hypothesize that the continuum originates from characteristic $\gamma$-rays scattering with environmental substances and depositing part of their energies.}, accounting for the majority of the flux. The aggregate flux of $\gamma$-rays between 73 to 3033~keV is $3.3\times 10^{7}$ (m$^2$$\cdot$sr$\cdot$hour)$^{-1}$. 

Other background sources, such as cosmic rays and radioactive contamination of the CZT crystal, could contribute to the measured spectrum. To inspect their contribution, we have placed the CZT detector into a lead box of a minimal thickness of 5 cm to obtain the spectrum with most of the ambient $\gamma$-rays shielded. The result demonstrates that other background sources only accounts for less than 1.6\% of the total counting rates, but in high energy regions above 2614.5~keV the shielded spectrum still has similar counting rates to the unshieled one, implying they are no longer dominated by environmental $\gamma$-rays\footnote{Based on calculations with empirical data of sea-level muon flux\cite{AUTRAN201877}, the muons may contribute most of the events above 2614.5 keV.}, as the rightmost bins in Fig.~\ref{fig:unfoldedspectra} indicates.

\subsection{Uncertainties estimation}

The uncertainties associated with the measured spectrum, as well as the construction of the migration matrix, are propagated to the unfolded spectrum. Also, the influence of ignoring the angle distribution of the $\gamma$ background should  be considered. In this section, uncertainty estimations involving several aspects are given in detail.

Regarding the statistical uncertainty of the migration matrix, we implement the estimation by sampling more migration matrices based on the original one and repeating the unfolding procedure. Assuming that the elements of the migration matrix before normalization follows a Gaussian distribution with $Mean = N_{ij}$ and $\sigma = \sqrt{N _{ij}}$, $10^3$ matrix samples are obtained by sampling this Gaussian distribution in each element of the original migration matrix, namely $N_{ij}^{new} \sim Gaus(E = N_{ij},\ \sigma = \sqrt{N_{ij}})$. Then, the unfolding procedure is repeated with different sampled migration matrices, while other factors remain unchanged. As a result, we calculate the Root Mean Square (RMS) values of each bin and the total flux respectively within the $10^3$ unfolded spectra, and take them as the  uncertainty limits.

A similar approach is utilized to estimate the uncertainty from measured spectrum. We sample $10^3$ different input spectra according to Gaussian distributions in each bin: $N_{i}^{new}\sim Gaus(E = N_{i},\ \sigma = \sqrt{N_{i}})$. Analogously, the unfolding is performed with the sampled spectra, from which we obtain the uncertainty limits of measured counts in the same approach as above.

For the systematic uncertainty caused by peak parameterization, we re-fit the linear correlation between peak-fitting parameters and peak energy using each combination of 3 (out of 4) calibrating data points, so as to find the max and min values of each linear function's gradient and intercept. Subsequently, $10^3$ new linear correlations of the three parameters ($A,B,\sigma$) are sampled by uniformly choosing gradient and intercept in the range given by re-fitting. These sampled linear functions are taken as an alternative of the original 4-points fitting to construct the migration matrices and repeat the unfolding procedure. As the uncertainties are asymmetric, the RMS values of the positive/negative (compared to the original one) unfolded results are calculated respectively and serve as uncertainty bars. 

In addition, the uncertainties associated with the detector model, such as the position of different components and the physics lists, are estimated by adjusting the corresponding settings. The resulting relative uncertainty is around $1.0\%$. The uncertainty from other background sources is also considered, as described in section 4.4, accounting for the major influence on the spectrum above 2614.5 keV.

Lastly, to estimate the impact of ignoring the angle distribution of $\gamma$ background, we equip the CZT detector with a lead collimator to measure the counting rate of $\gamma$-rays from different directions. After a background subtraction with the fully shielded data, the result illustrates a difference of $\pm 25.7\%$ among four directions. Without significant change in spectrum shapes observed, the uncertainty is directly assigned to both the total flux and each bin.

Assuming that the contributions from different sources are independent, the total uncertainty is given by their sum in quadrature. The uncertainty bar in each bin is shown in Fig.~\ref{fig:unfoldedspectra}, and the uncertainties on the total flux are listed in Table~\ref{tab:uncertaintyTotal}. The isotropic approximation introduces the majority of the uncertainties, which possibly comes from the anisotropy of the $\gamma$ sources and absorbing materials (e.g. building materials). In addition, peak parameterization uncertainties can cause counts migration between adjacent bins and contribute significantly to the bins around the peaks.

\begin{table}[!htbp]
\small
\caption{\label{tab:uncertaintyTotal}Uncertainty budget of the aggregate $\gamma$ flux.}
\begin{center}
\begin{tabularx}{0.45\textwidth}{ccc}
\hline
\multirow{2}{*}{Uncertainty-related item}&\multicolumn{2}{c}{Uncertainty Value \%}\\ \cline{2-3}
&\multicolumn{1}{c}{Positive}&\multicolumn{1}{c}{Negative} \\
\hline
Migration matrix&0.02&-0.02\\
Measured events&0.06&-0.06\\
Peak parameterization&0.44&-0.59\\
Geant4 modeling&1.03&-1.03\\
Other background&0.00&-1.57\\
Assumption of isotropy&25.74&-25.74\\
\hline
Total&25.76&-25.81\\
\hline
\end{tabularx}
\end{center}
\end{table}


\section{Utilizing and validation of the unfolded spectrum}\label{sec:design}
\begin{figure}[!htbp]
\begin{center}
\subfigure[]{
\includegraphics
  [width=6cm]
  {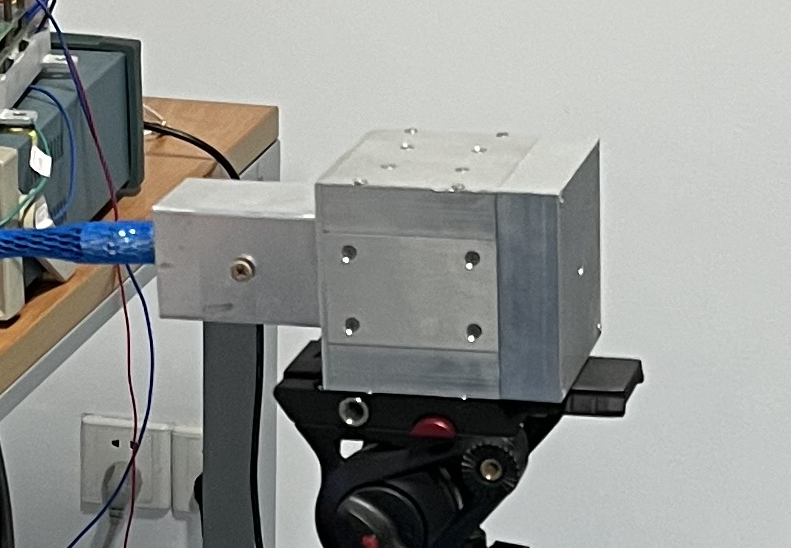}}
\subfigure[]{
\includegraphics
  [width=7cm]
  {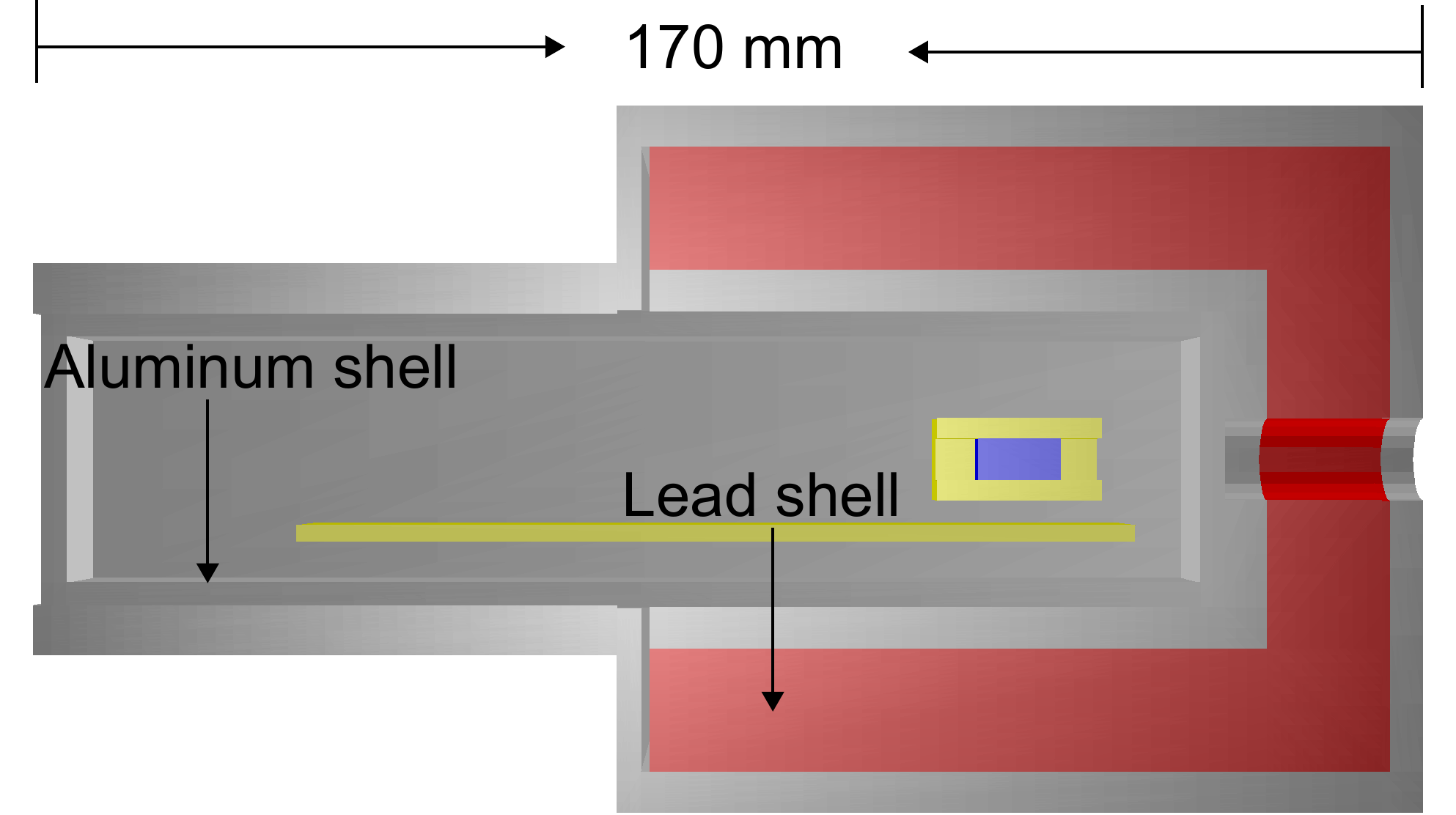}}
\caption{\label{fig:leadshell}(a) Experimental setup. (b) A section view of GEANT4 geometry of the detector with lead shell.}
\end{center}
\end{figure}

Utilizing the reconstructed environmental $\gamma$ flux and spectrum, the intensity and spectrum of $\gamma$-ray background under different conditions of shielding setup can be evaluated. 
In this section, we conduct measurement and simulation using the CZT detector with a lead shell as $\gamma$ shielding, so as to change the response of the detection system to validate the unfolded result and give an instance of application. The structure of the lead shell mainly consists of a semi-closed lead block with a thickness of 15 mm, an inner aluminum holder of 6 mm in thickness, a outer aluminum shell of 5 mm in thickness, and a hole in the front of 3 mm in radius, as shown in Fig.~\ref{fig:leadshell}. The configuration of particle source is the same as that Fig.~\ref{fig:CZTsimu} depicts, and the exposure time in simulations is derived from Eq.~\ref{eq:flux} with the number of generated events, the value of aggregate flux, and the geometric acceptance of the spherical particle source already known.

\begin{figure}[!htbp]
\begin{center}
\includegraphics
  [width=10cm]
  {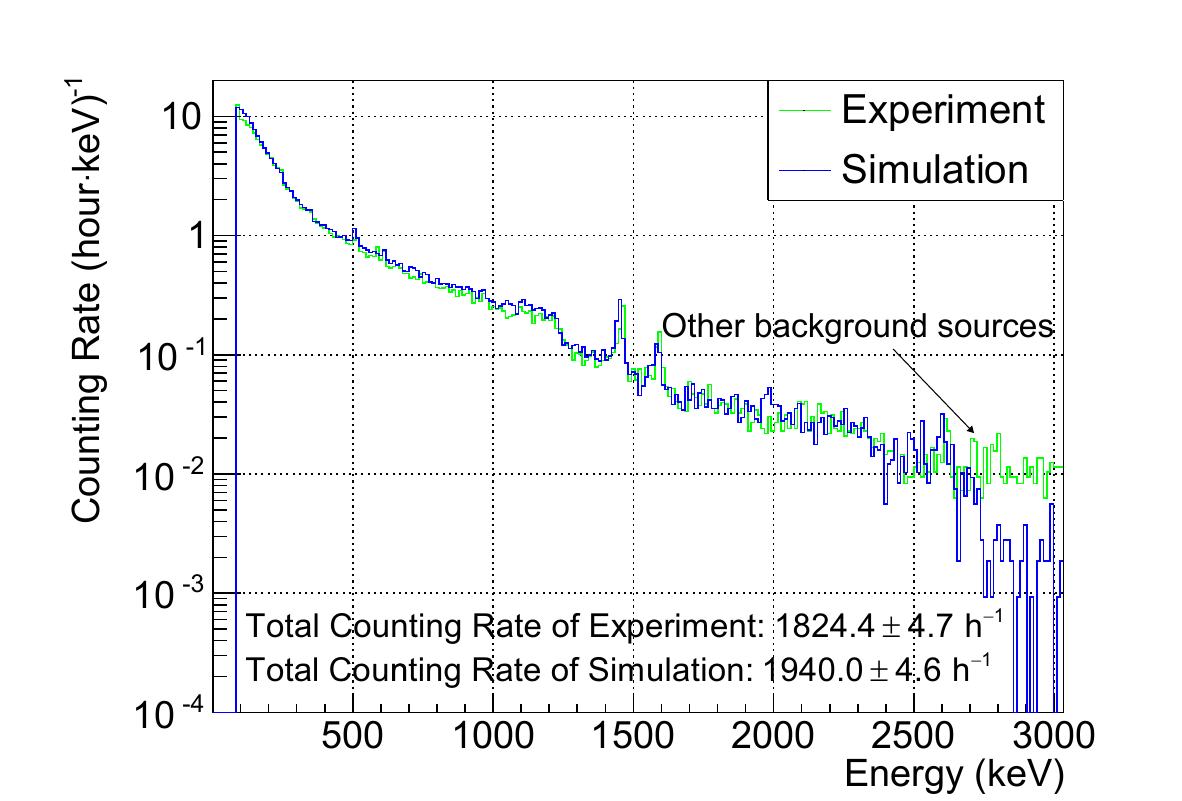}
\caption{\label{fig:validationResult}Comparison between simulated and measured $\gamma$ sepctrum of CZT detector with lead shell.}
\end{center}
\end{figure}

Figure~\ref{fig:validationResult} illustrates the comparison of $\gamma$-rays spectra between measurement and the corresponding MC simulation. The spectra are in good consistency within the range below $2614.5$~keV, with the difference in counting rate around $6.3 \%$. As $\gamma$ radiation from radioactive nuclides ends at $2614.5$~keV, the response model is not suitable for spectrum above this energy induced by other radiation types, which especially leads to the differences in high energy regions.

As the validation demonstrates, the method provides a good evaluation of $\gamma$ background intensity, including the counting rates both in aggregate and in given energy regions, which serves as an excellent fundation of background estimation. Simulations of other background sources can also be incorporated in the current model for further optimization of shielding construction and background reduction~\cite{CUORE:2017ztm,4437209}.

\section{Summary}

In this paper, we present a detailed procedure to reconstruct environmental $\gamma$ flux spectra and evaluate the $\gamma$ background intensity. We have calibrated a CZT detector, modeled it in MC simulations and used it for $\gamma$ spectrum measurement. After gaining a comprehensive description of the CZT detector via MC implementation, we are able to deconvolute the measured spectrum applying iterative Bayesian unfolding. Lastly, we have illustrated an evaluation of $\gamma$ background rate with the unfolded environmental $\gamma$ flux spectrum to validate the effectiveness of this approach.

The measurement around our above-ground platform has revealed an aggregate $\gamma$ flux of $3.3 \pm 0.9\times 10^{7}$~(m$^2$$\cdot$sr$\cdot$hour)$^{-1}$ from 73 to 3033~keV, as well as the detailed charateristic peaks in the unfolded spectrum mainly attributed to $^{232}$Th series, $^{238}$U series and $^{40}$K. 
Since $\gamma$ background is an important concern of various low-background experiments, the transferability and high feasibility of the approach presented in this paper makes it promising for background reduction in the design and construction of different experiments or laboratories.

\section{Acknowledgment}

This work was supported by National Key R\&D Program of China (2023YFA1607203), National Natural Science Foundation of China (12005225, 12141505) and the Fundamental Research Funds for the Central Universities, China (WK2360000015). We are also grateful to Doug Pinckney from Massachusetts Institute of Technology and Ziqing Hong from University of Toronto for valuable discussions and suggestions.






\begin{thebibliography}{10}
\expandafter\ifx\csname url\endcsname\relax
  \def\url#1{\texttt{#1}}\fi
\expandafter\ifx\csname urlprefix\endcsname\relax\def\urlprefix{URL }\fi
\expandafter\ifx\csname href\endcsname\relax
  \def\href#1#2{#2} \def\path#1{#1}\fi

\bibitem{CUORE:2021mvw} CUORE Collaboration, Search for Majorana neutrinos exploiting millikelvin cryogenics with CUORE, Nature, 604 (7904) (2022) 53-58.

\bibitem{augier2022final} C. Augier, A.S. Barabash, F. Bellini, et al., Final results on the $0\nu\beta\beta$ decay half-life limit of $^{100}$Mo from the CUPID-Mo experiment, The European Physical Journal C, 82 (11) (2022) 1-20.

\bibitem{ahmine2023test} A. Ahmine, I.C. Bandac, A.S. Barabash, et al., Test of $^{116}$CdWO$_4$ and Li$_2$MoO$_4$ scintillating bolometers in the CROSS underground facility with upgraded detector suspension, Journal of Instrumentation, 18 (12) (2023) P12004.

\bibitem{alkhatib2021light} I. Alkhatib, D.W.P. Amaral, T. Aralis, et al., Light dark matter search with a high-resolution athermal phonon detector operated above ground, Physical Review Letters, 127 (6) (2021) 061801.


\bibitem{thome2017repair} C. Thome, S. Tharmalingam, J. Pirkkanen, A. Zarnke, T. Laframboise, D.R. Boreham, et al., The repair project: examining the biological impacts of sub-background radiation exposure within SNOLAB, a deep underground laboratory, Radiation Research, 188 (4.2) (2017) 470-474.


\bibitem{Vepsalainen:2020trd} A. Vepsäläinen, et al., Impact of ionizing radiation on superconducting qubit coherence, Nature, 584 (7822) (2020) 551-556.


\bibitem{CUORE:2017ztm} CUORE Collaboration, The projected background for the CUORE experiment, The European Physical Journal C, 77 (8) (2017) 543. 

\bibitem{eisen1998cdte} Y. Eisen, A. Shor, CdTe and CdZnTe materials for room-temperature x-ray and gamma ray detectors, Journal of Crystal Growth, 184 (1998) 1302-1312.

\bibitem{mortreau2001characterisation} P. Mortreau, R. Berndt, Characterisation of cadmium zinc telluride detector spectra--application to the analysis of spent fuel spectra, Nuclear Instruments and Methods in Physics Research Section A: Accelerators, Spectrometers, Detectors and Associated Equipment, 458 (1-2) (2001) 183-188.


\bibitem{namboodiri1996gamma} M. Namboodiri, A. Lavietes, J. McQuaid, Gamma-ray peak shapes from cadmium zinc telluride detectors, Lawrence Livermore National Lab.(LLNL), Livermore, CA (United States), Report, 1996.

\bibitem{dardenne1999cadmium} Y. Dardenne, T. Wang, A. Lavietes, G. Mauger, W. Ruhter, S. Kreek, Cadmium zinc telluride spectral modeling, Nuclear Instruments and Methods in Physics Research Section A: Accelerators, Spectrometers, Detectors and Associated Equipment, 422 (1-3) (1999) 159-163.

\bibitem{Brun:1997pa} R. Brun, F. Rademakers, ROOT: An object oriented data analysis framework, Nuclear Instruments and Methods in Physics Research Section A: Accelerators, Spectrometers, Detectors and Associated Equipment, 389 (1997) 81-86.

\bibitem{GEANT4:2002zbu} S. Agostinelli, et al., GEANT4--a simulation toolkit, Nuclear Instruments and Methods in Physics Research Section A: Accelerators, Spectrometers, Detectors and Associated Equipment, 506 (2003) 250-303.



\bibitem{geant4EM} Geant4 Collaboration, Gamma Incident, available at \url{https://geant4-userdoc.web.cern.ch/UsersGuides/PhysicsReferenceManual/html/electromagnetic/gamma_incident/index.html} (accessed in February, 2024).

\bibitem{geant4EM2} Geant4 Collaboration, Electron Incident, available at \url{https://geant4-userdoc.web.cern.ch/UsersGuides/PhysicsReferenceManual/html/electromagnetic/electron_incident/index.html} (accessed in February, 2024).

\bibitem{joel2018monte} C. S. Joel, M. N. Moyo, J. N. M. Eric, O. Motapon, D. Strivay, Monte Carlo method for gamma spectrometry based on GEANT4 toolkit: Efficiency calibration of BE6530 detector, Journal of Environmental Radioactivity, 189 (2018) 109-119.

\bibitem{liu2017detection} C. Liu, K. Ungar, D. Pierce, I. Hoffman, W. Zhang, Detection efficiency calculations using Geant4 for a broad-energy germanium gamma spectrometer, Journal of Radioanalytical and Nuclear Chemistry, 312 (2017) 471-478.

\bibitem{Niu_2015} S.-L. Niu, X. Cai, Z.-Z. Wu, et al., Simulation of background reduction and Compton suppression in a low-background HPGe spectrometer at a surface laboratory, Chinese Physics C, 39 (8) (2015) 086002.

\bibitem{DAgostini:1994fjx} G. D'Agostini, A Multidimensional unfolding method based on Bayes' theorem, Nuclear Instruments and Methods in Physics Research Section A: Accelerators, Spectrometers, Detectors and Associated Equipment, 362 (1995) 487-498.

\bibitem{AUTRAN201877} J.L. Autran, D. Munteanu, T. Saad Saoud, S. Moindjie, Characterization of atmospheric muons at sea level using a cosmic ray telescope, Nuclear Instruments and Methods in Physics Research Section A: Accelerators, Spectrometers, Detectors and Associated Equipment, 903 (2018) 77-84.

\bibitem{Bueno:2022dva} E. F. Bueno, F. Barão, M. Vecchi, Iterative-Bayesian unfolding of cosmic-ray isotope fluxes measured by AMS-02, Nuclear Instruments and Methods in Physics Research Section A: Accelerators, Spectrometers, Detectors and Associated Equipment, 1046 (2023) 167695.

\bibitem{Brenner:2019lmf} L. Brenner, R. Balasubramanian, C. Burgard, et al., Comparison of unfolding methods using RooFitUnfold, International Journal of Modern Physics A, 35 (24) (2020) 2050145.

\bibitem{4437209} C. Hagmann, D. Lange, D. Wright, Cosmic-ray shower generator (CRY) for Monte Carlo transport codes, 2007 IEEE Nuclear Science Symposium Conference Record, 2 (2007) 1143-1146.





\end{thebibliography}
\end{document}